\documentclass[aps,prl,twocolumn,floatfix,showpacs,citeautoscript,longbibliography]{revtex4-1}

\usepackage{pdfpages}

\usepackage{amsmath}
\usepackage{amssymb}
\usepackage{amsfonts}
\usepackage{graphicx,graphics}

\newcommand{\ket}[1]{\lvert #1\rangle}
\newcommand{\bra}[1]{\langle#1 \rvert}
\newcommand{\abs}[1]{\lvert #1 \rvert}

\begin{document}

\title{Correlated Coulomb drag in capacitively coupled quantum-dot structures}
\author{Kristen Kaasbjerg}
\email{cosby@fys.ku.dk}
\author{Antti-Pekka Jauho}
\affiliation{Center for Nanostructured Graphene (CNG), Department of Micro- and
  Nanotechnology, Technical University of Denmark, DK-2800 Kgs. Lyngby, Denmark}
\date{\today}

\begin{abstract}
  We study theoretically Coulomb drag in capacitively coupled quantum dots
  (CQDs) --- a bias-driven dot coupled to an unbiased dot where transport is due
  to Coulomb mediated energy transfer drag. To this end, we introduce a
  master-equation approach that accounts for higher-order tunneling
  (cotunneling) processes as well as energy-dependent lead couplings, and
  identify a mesoscopic Coulomb drag mechanism driven by \emph{nonlocal}
  multielectron cotunneling processes. Our theory establishes the conditions
  for a nonzero drag as well as the direction of the drag current in terms of
  microscopic system parameters. Interestingly, the direction of the drag
  current is \emph{not} determined by the drive current, but by an interplay
  between the energy-dependent lead couplings. Studying the drag mechanism in a
  graphene-based CQD heterostructure, we show that the predictions of our theory
  are consistent with recent experiments on Coulomb drag in CQD systems.
\end{abstract}

\pacs{73.23.-b, 72.80.Vp, 73.23.Hk, 73.63.Kv}
\maketitle

Electronic systems brought into close proximity may exhibit Coulomb
drag~\cite{Rojo:ElectronDrag,Levchenko:CoulombDrag}: a current in one system
induces a current (or a voltage) in a nearby undriven system. Importantly, the
effect arises solely due to Coulomb interactions between the charge carriers in
the two systems. Coulomb drag has been studied extensively in bulk
two-dimensional systems, both
experimentally~\cite{Tulipe:New,West:Mutual,Shtrikman:Coupled} and
theoretically~\cite{Smith:CoulombDrag,Hu:CoulombDrag,Kinaret:Linear,Oreg:CoulombDrag},
and has recently experienced a revival in one-dimensional
systems~\cite{Jauho:Mesoscopic,Tarucha:Negative,Reno:Positive,Reno:1DCoulombDrag,Jauho:PlasmonMediated}
and graphene
heterostructures~\cite{Sarma:Theory,Sarma:CoulombDrag,Tutuc:Coulomb,Katsnelson:CoulombDrag,Ponomarenko:Strong,Levitov:Coulomb}.

In mesoscopic systems with broken translational invariance, e.g. quantum point
contacts or quantum dots (QDs), momentum is not a good quantum number as in
extended systems. Instead of momentum transfer, it is more natural to view
\emph{mesoscopic} Coulomb
drag~\cite{Jauho:Coulomb,Oppen:Coulomb,Wang:Correlated,Kamenev:Coulomb,Buttiker:Mesoscopic}
as an interaction mediated energy transfer between the drive and the drag
system. Such energy-transfer drag plays a central role in, for example, quantum
measurements where a detector and a system exchange energy in a measurement on
the system~\cite{Schoelkopt:Introduction}. In this case, the drag can either
constitute the signal in the detector generated by the measured quantum noise in
the system~\cite{Kouwenhoven:DQD,Kouwenhoven:Using,Gossard:Frequency}, or be a
disturbance in the system due to the
measurement~\cite{Wegscheider:DQD,Ludwig:Phonon}, i.e. detector backaction.

In addition to energy transfer, Coulomb drag in capacitively coupled QDs (CQDs)
relies on an asymmetry in the drag system~\cite{Buttiker:Mesoscopic}. This has
been demonstrated in coupled double quantum dots~\cite{Fujisawa:Bidirectional},
and recently in coupled single QDs~\cite{Ensslin:Measurement,Stampfer:Back}
where the asymmetry originates from the couplings to the leads. In the latter,
Coulomb-drag effects beyond conventional mesoscopic QD
drag~\cite{Buttiker:Mesoscopic} were reported~\cite{Ensslin:Measurement}. Not
only are such effects of fundamental scientific interest, but they may also be
important for the performance of thermoelectric CQD
devices~\cite{Buttiker:Optimal,Jordan:Thermoelectric,Molenkamp:Three,Pekola:Onchip}.

\begin{figure}[!b]
  \centering
  % \includegraphics[width=0.23\linewidth]{stability}
  % \hspace{1cm}
  % \includegraphics[width=0.55\linewidth]{heterostructure}
  % \includegraphics[width=0.6\linewidth]{processes}
  % \includegraphics[width=0.93\linewidth]{fig1}
  \includegraphics[width=1.0\linewidth]{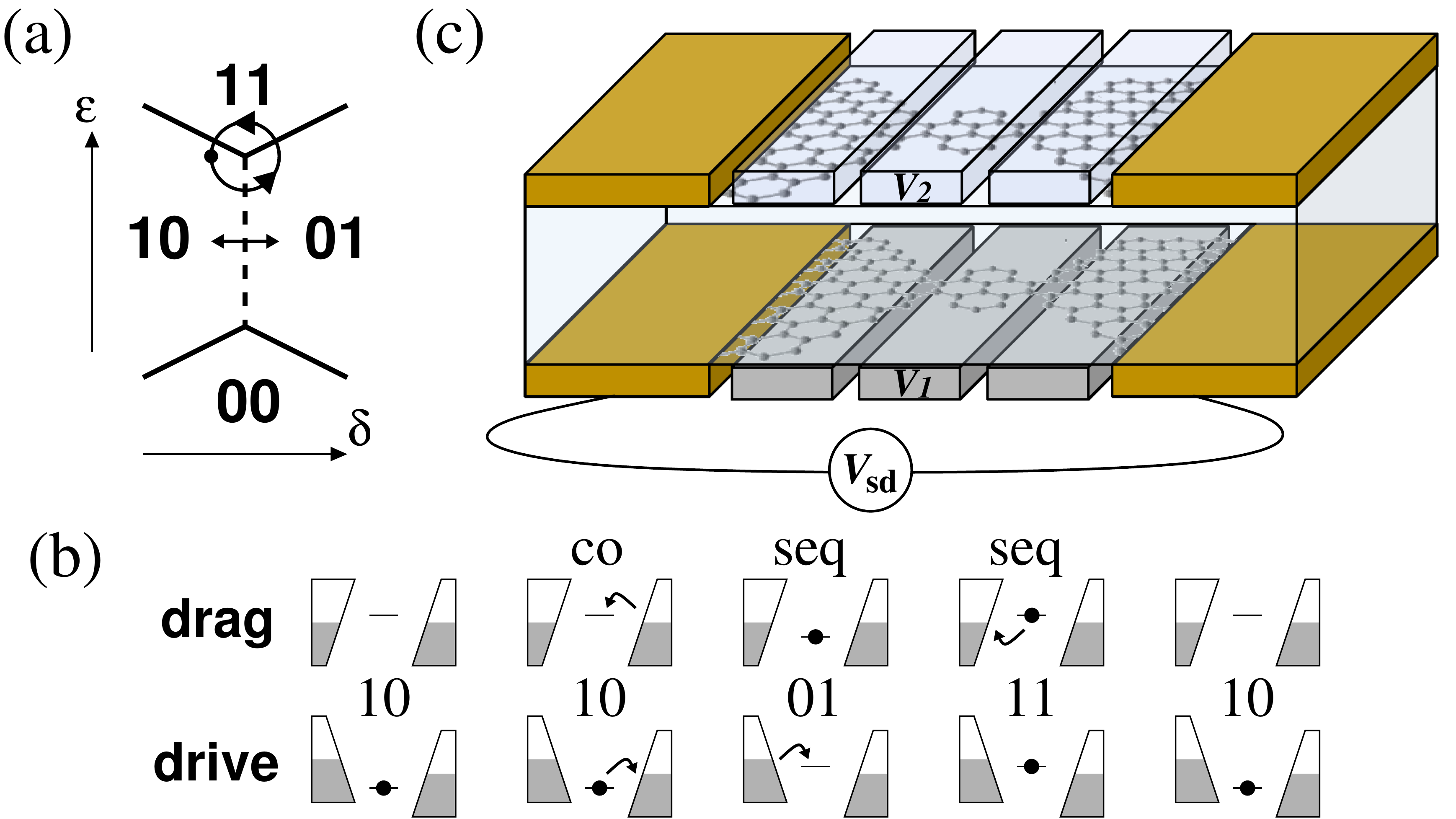}
  \caption{(a) Charge stability diagram of two capacitively coupled QDs as a
    function of their gate detuning $\delta=V_2 - V_1$ and common gate
    $\epsilon=V_1 + V_2$. (b) Sequence of sequential and cotunneling processes
    underlying the drag mechanism in the vicinity of the triple points [closed
    circle in (a)]. Away from the triple points, the drag is driven by
    cotunneling only [arrow in (a)]. Energy-dependent lead couplings are
    essential for the mechanism to induce a directional current in the drag
    system. (c) Illustration of a graphene-based CQD heterostructure with two
    QDs defined in stacked graphene layers separated by a thin isolating
    dielectric~\cite{Ensslin:Measurement}. A series of top and bottom gates
    control the potentials on the quantum dots ($V_{1/2}$) and their adjacent
    graphene leads.}
\label{fig:overview}
\end{figure}
In this work we introduce a theoretical framework for the description of Coulomb
drag in CQDs taking into account higher-order tunneling (cotunneling) processes,
and thereby going beyond conventional QD drag~\cite{Buttiker:Mesoscopic}. We
uncover a drag mechanism driven by \emph{nonlocal} correlated multielectron
cotunneling processes where energy transfer is mediated by bias-induced
switching of the CQD states. At the triple points of the CQD charge stability
diagram~\cite{Kouwenhoven:RMP} sketched in Fig.~\ref{fig:overview}(a), it
resembles a stochastic ratchet mechanism which, like charge pumping
mechanisms~\cite{Janssen:Gigahertz}, results in drag via excursions [in state
space; see Fig.~\ref{fig:overview}(b)] around the triple points. Our theory
pinpoints the conditions for drag in terms of microscopic quantities and shows
that the direction of the drag current is independent on the drive current and
exhibits a nontrivial dependence on the lead couplings in the drag system.

We demonstrate the rich properties of the drag mechanism by studying drag in the
graphene-based CQD structure illustrated in Fig.~\ref{fig:overview}(c). Such
experimentally realizable graphene-based QD structures are unique due their
large
tunability~\cite{Geim:Chaotic,Ensslin:Tunable,Jiang:Graphene,Ensslin:Transport,Ensslin:Localized},
large interdot charging energies~\cite{Ensslin:Measurement}, and built-in
graphene leads. We envision structures in which local gating allows to control
the chemical potentials of the lead
regions~\cite{Gordon:Transport,Kim:Electronic} and create, e.g., $p$-QD-$n$
junctions across the individual QDs. As we demonstrate below, this opens the
opportunity to control the direction of the drag current. Finally, we elaborate
on the role of the drag mechanism in the recently observed Coulomb drag in a
graphene-based CQD heterostructure~\cite{Ensslin:Measurement}.

\textbf{\emph{General model and theory.}}---We consider a generic (spinless)
model for two capacitively coupled QDs---a \emph{biased} drive ($i=1$) and an
\emph{unbiased} drag ($i=2$) QD---with one level each,
$H_\text{CQD}= \sum_i \varepsilon_i n_i + U_{12} n_1n_2$, where the dot levels
are controlled by gate voltages $\varepsilon_i = - eV_i$,
$n_i=d_i^\dagger d_i^{\phantom\dagger}$ is the dot occupation, and
$U_{12}=e^2/2C$ is the capacitive inter-dot Coulomb interaction. The dots are
coupled to separate sets of source and drain contacts,
$H_\alpha = \sum_k \xi_{\alpha k} c_{\alpha k}^\dagger c_{\alpha
  k}^{\phantom\dagger}$,
$\xi_{\alpha k} = \varepsilon_k - \mu_\alpha$ ($\alpha=L_i,R_i$;
$\mu_{L_1/R_1}=\pm eV_\text{sd}/2+\mu_0$ and $\mu_{L_2/R_2}=\mu_0$), via tunnel
Hamiltonians
$H_T = \sum_{\alpha k} t_{\alpha k} c_{\alpha k}^\dagger d_i^{\phantom\dagger} +
\text{hc}$.
In contrast to the usual wide-band approximation where the lead couplings are
assumed constant, we here consider energy-dependent couplings
$\Gamma_\alpha(\varepsilon) = 2\pi \rho_\alpha(\varepsilon)
\abs{t_\alpha(\varepsilon)}^2$,
where $\rho_\alpha$ is the density of states (DOS) in lead $\alpha$ and
$t_\alpha$ is the tunnel coupling. Like in conventional QD
drag~\cite{Buttiker:Mesoscopic,Ensslin:Measurement}, this is the key ingredient
for the drag mechanism described below.

We describe the transport through the drive and drag dots with a master equation
approach valid for $k_BT\gtrsim \Gamma_\alpha$~\cite{Flensberg}. The occupation
probabilities $p_m$ for the CQD states,
$\ket{m} = \ket{n_1 n_2} \in \{\ket{00},\ket{10},\ket{01},\ket{11}\}$, are
determined by the rate equations
\begin{align}
  \label{eq:rateequation}
  \dot{p}_m & = - p_m \sum_{n\neq m} \Gamma_{m n} 
                  + \sum_{n\neq m} p_n \Gamma_{n m} ,
\end{align}
which together with the normalization condition $\sum_m p_m=1$ are
solved for the steady-state probabilities, i.e. $\dot{p}_m=0$.

The rates for tunneling-induced transition between the states are obtained from
the generalized Fermi golden rule~\cite{Flensberg},
\begin{equation}
  \label{eq:goldenrule}
  \Gamma_{mn} = \frac{2\pi}{\hbar} \sum_{i'f'}
      W_{i'} \abs{ \bra{f} T \ket{i} }^2 \delta(E_f - E_i) .
\end{equation}
Here, $\ket{i/f}=\ket{m/n}\otimes\ket{i'/f'}$ are products of QD and lead
states, the sum is over possible initial $\ket{i'}$ and final $\ket{f'}$ states
of the leads, $W_{i'}$ is the probability for the initial lead state $\ket{i'}$,
and $T = H_T + H_T G_0 H_T + \ldots$ is the $T$ matrix with
$G_0=\tfrac{1}{E_i - H_0}$ denoting the Green function in the absence of
tunneling, i.e. $H_0=H_\text{CQD} + \sum_\alpha H_\alpha$. The correlations
between the occupations of the QDs are fully accounted for in $G_0$ which is
treated exactly.

To lowest order in the tunneling Hamiltonian, the transitions between the states
are given by sequential tunneling processes with rates
\begin{align}
  \Gamma_{m , 11}^{\alpha} & = \hbar^{-1}\Gamma_\alpha(\Delta_{m,11}) 
      f_\alpha(\Delta_{m,11})    \\
  \Gamma_{m , 00}^{\alpha} & = \hbar^{-1}\Gamma_\alpha (\Delta_{00,m}) 
      \left[ 1- f_\alpha(\Delta_{00,m}) \right]  \\
  \Gamma_{00 , n}^{\alpha} & = \hbar^{-1}\Gamma_\alpha(\Delta_{00,n}) 
      f_\alpha(\Delta_{00,n})    \\
  \Gamma_{11 , n}^{\alpha} & = \hbar^{-1}\Gamma_\alpha (\Delta_{n, 11})
      \left[ 1- f_\alpha(\Delta_{n,11}) \right] ,
\end{align}
where $m,n\in \{10,01\}$, $f_\alpha$ is the Fermi function in lead $\alpha$, and
$\Delta_{mn} = E_n - E_m$.

The next-to-leading order term in the $T$ matrix gives rise to \emph{elastic}
and \emph{inelastic} cotunneling through the individual
QDs~\cite{Nazarov:Virtual,Matveev:Theory,Kouwenhoven:Electron}. In addition, we
identify a \emph{nonlocal} cotunneling process mediated by the capacitive
interdot coupling. This is a correlated two-electron tunneling event in which
the CQD switches between the $10 \leftrightarrow 01$ states in one coherent
process. The rate for nonlocal cotunneling processes which transfer an electron
from lead $\alpha$ to lead $\beta$ is given by
\begin{align}
  \Gamma_{mn}^{\alpha\beta} 
      & = 
        \int\! \frac{d\varepsilon}{2\pi\hbar} \,
        \Gamma_\alpha(\varepsilon + \Delta_{mn}) \Gamma_\beta(\varepsilon)
        f_\alpha(\varepsilon + \Delta_{mn}) [ 1 - f_\beta(\varepsilon) ]
        \nonumber \\
      & \quad \times \left\vert
          \frac{1}{\varepsilon + \Delta_{11,n} } -
          \frac{1}{\varepsilon + \Delta_{m,00}}
        \right\vert^2 , 
\end{align}
where $m,n\in \{10,01\}$ and the terms in the last line account for the energy
of the virtually occupied intermediate $00$/$11$ states. To evaluate the
cotunneling rates at finite temperature and bias, we have generalized the
commonly applied regularization
scheme~\cite{Matveev:Cotunneling,Sela:Thermopower} to the situation with
energy-dependent lead couplings~\cite{supplementary}.

From the solution to the master equation~\eqref{eq:rateequation}, the currents
in the various leads are obtained as
\begin{equation}
  \label{eq:current}
    I_{\alpha} = -e \sum_{mn} p_m  \left( 
        \Gamma^{\rightarrow \alpha}_{mn} -
        \Gamma^{\alpha \rightarrow}_{mn}
      \right) ,
\end{equation}
where $\Gamma^{\rightarrow \alpha}$ ($\Gamma^{\alpha\rightarrow}$) denotes the
rate for processes that transfer an electron into (out of) lead $\alpha$, and
the drive and drag currents are defined as $I_\text{drive} = I_{L_1} = -I_{R_1}$
and $I_\text{drag} = I_{L_2} = -I_{R_2}$, respectively.

\textbf{\emph{Drag mechanism.}}---In the following, we focus on the regime of
low bias on the drive QD, $eV_\text{sd} \lesssim U_{12}$, where the conventional
drag mechanism~\cite{Buttiker:Mesoscopic} is suppressed. Fixing the gate
voltages to, e.g., the point below the 10,11 degeneracy line at the upper triple
point in Fig.~\ref{fig:overview}(a), a finite bias on the drive QD opens for the
sequence of transitions illustrated in Fig.~\ref{fig:overview}(b),
\begin{equation}
  \label{eq:sequence}
  \ket{10} \quad \overset{\text{co}}{\leftrightarrow}  \quad 
  \ket{01} \quad \overset{\text{seq}}{\leftrightarrow} \quad 
  \ket{11} \quad \overset{\text{seq}}\rightarrow   \quad \ket{10} .
\end{equation}
For $eV_\text{sd}>\abs{\Delta_{10,01}},\abs{\Delta_{01,11}} \gg k_BT$, the two
first transitions are open in both directions, whereas the third transition is
only open in the forward direction because the drag QD is unbiased. In addition
to a drive current, this may induce a drag current via steps where the drag QD
is repeatedly filled and emptied. This is possible via the first step alone
(cotunneling-only), or through the full sequence (cotunneling-assisted
drag). The two mechanisms govern the drag, respectively, away from and at the
triple points [cf. Fig.~\ref{fig:overview}(a)]. Note that the \emph{nonlocal}
cotunneling process is instrumental in both cases.

In order to generate a drag current, the drag QD must be filled and emptied at
preferentially separate leads. This requires an asymmetry in the drag system. To
identify the exact conditions, we expand the lead couplings around the
equilibrium chemical potentials $\mu_{0}$,
$\Gamma_\alpha(\varepsilon) \approx \Gamma_{\alpha 0} + \xi \partial
\Gamma_\alpha$,
where $\xi = \varepsilon - \mu_0$, $\Gamma_{\alpha 0} = \Gamma_\alpha(\mu_0)$,
and
$\partial\Gamma_\alpha=\partial\Gamma_\alpha/\partial\varepsilon\vert_{\varepsilon=\mu_0}$.
Along the 10,01 degeneracy line where $\Delta_{10,01}=0$, and in the nonlinear
regime $eV_\text{sd}\gg k_BT$ (but still $eV_\text{sd} < U_{12}$) where the
transport in the drive QD is \emph{unidirectional}, we find for the drag
current,
\begin{equation}
  \label{eq:drag_triple}
  I_\text{drag} \sim \frac{\Gamma_{L_1 0}\Gamma_{R_1 0}
    (\Gamma_{L_2 0}\partial\Gamma_{R_2 0} - \Gamma_{R_2 0}\partial\Gamma_{L_2 0})}
    {\Gamma_{L_2 0} + \Gamma_{R_2 0}}  F(V_\text{sd} ) ,
\end{equation}
where $F(V_\text{sd})=V_\text{sd}^2,\, \log V_\text{sd}$ for cotunneling-only
and cotunneling-assisted drag, respectively. The factor in parentheses in the
numerator gives the conditions for drag. Notably, the drag is zero if the lead
couplings to the drag QD are constant or differ by a multiplicative
factor. Furthermore, the direction of the drag current is determined by two
factors concerning the lead couplings to the drag QD: (i) their asymmetry, and
(ii) their derivatives.

\textbf{\emph{Drag in graphene-based CQDs.}}---We now proceed to study the drag
effect in an idealized version of the graphene-based CQD structure illustrated
in Fig~\ref{fig:overview}(c). The QDs are assumed to be connected to bulk
graphene leads with linear DOS,
$\rho_\alpha(\varepsilon) = \frac{g_s g_v}{2\pi (\hbar v_F)^2}
\abs{\varepsilon-E_{\alpha0}}$,
which govern the energy dependence of the lead couplings, i.e.
$\Gamma_\alpha(\varepsilon) =2\pi \rho_\alpha(\varepsilon)\abs{t_\alpha}^2$
where $t_\alpha$ is constant, and where the positions of the Dirac points,
$E_{\alpha0}=-eV_{\alpha}$, are controlled by local gates [see
Fig.~\ref{fig:drag_sign}(a)]. This allows to tune both the strength of the lead
couplings, $\Gamma_{\alpha 0} \propto \abs{\mu_{0}-E_{\alpha0}}$, as well as
their derivatives, $\partial \Gamma_\alpha \gtrless 0$ on the upper/lower Dirac
cones. In order to meet the conditions for a nonzero drag current,
$E_{L_2 0} \neq E_{R_2 0}$ like in Fig.~\ref{fig:drag_sign}(a) is
necessary. Asymmetric tunnel couplings alone,
$t_{L_2}\neq t_{R_2}\rightarrow \Gamma_{L_2}(\varepsilon) \propto
\Gamma_{R_2}(\varepsilon)$, is not enough.

\begin{figure}[!t]
  \centering
  \includegraphics[width=1.0\linewidth]{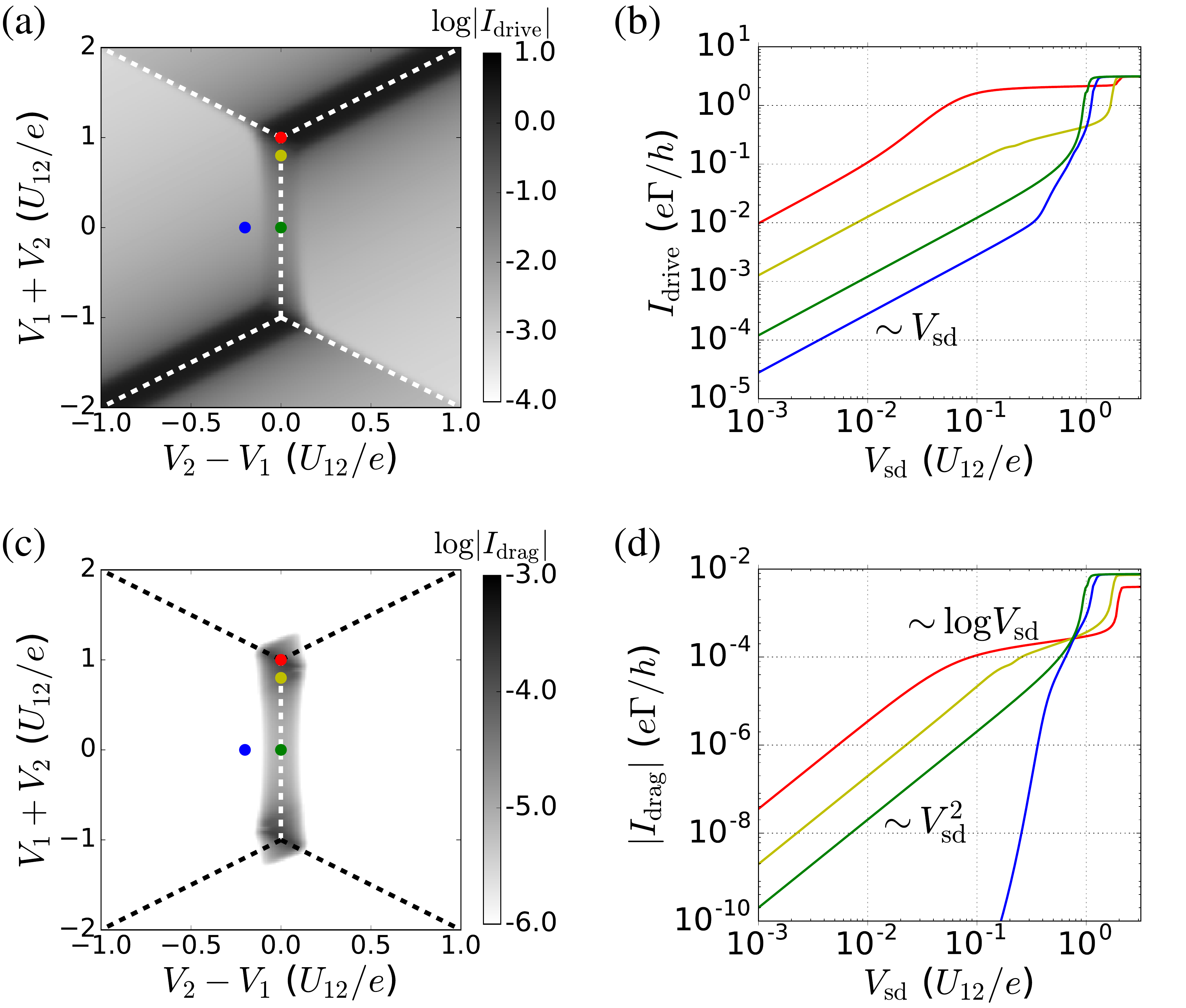}
  \caption{Drive (top) and drag current (bottom) for the graphene-based CQD in
    Fig.~\ref{fig:overview}(c), with the voltage configuration in
    Fig.~\ref{fig:drag_sign}(a). (a),(c) Current vs common gate and gate
    detuning with a bias voltage $eV_\text{sd}=0.2$ applied to the drive
    QD. (b),(d) Bias dependence of the drive and drag currents at the gate
    voltages $(V_2-V_1,V_1+V_2)$ marked by dots in the left plots [red:
    $(0.0,1.0)$, yellow: $(0.0,0.8)$, green: $(0.0,0.0)$, blue:
    $(-0.2,0.0)$]. Parameters (in units of $U_{12}$): $U_{12} = 1$,
    $\Gamma_{L_10/R_10} = \Gamma_{L_20/R_20} = 0.01\equiv\Gamma$,
    $\partial\Gamma_{L_2} = -\partial\Gamma_{R_2}$, $t_{L_2}=t_{R_2}$,
    $k_BT=0.01$.}
\label{fig:drag_vertex}
\end{figure}
In Figs.~\ref{fig:drag_vertex}(a) and~\ref{fig:drag_vertex}(c) we show the
numerically calculated currents through the drive and drag QDs as a function of
gate voltages for the situation in Fig.~\ref{fig:drag_sign}(a) and
$k_B T \ll eV_\text{sd} < U_{12}$. The current through the drive QD in
Fig.~\ref{fig:drag_vertex}(a) is nonzero along the $00,10$ and $01,11$
degeneracy lines, and the 10,01 degeneracy line where it is dominated by,
respectively, sequential tunneling and nonlocal cotunneling. In addition,
elastic cotunneling through the drive QD appears as a background in the
Coulomb-blockaded regions.

The induced drag current is shown in Fig.~\ref{fig:drag_vertex}(c). A finite
drag current is observed along the 10,01 degeneracy line where the nonlocal
cotunneling channel is open. With the bias applied symmetrically to the drive
dot, this is the case for
$e\abs{V_2-V_1} = \abs{\Delta_{10,01}} < eV_\text{sd}/2$. Away from the triple
points, $\abs{\Delta_{10/01,00/11}}\gg eV_\text{sd}$, the drag is driven by
nonlocal cotunneling only. In the vicinity of the upper (lower) triple point,
$\abs{\Delta_{01,11}} \lesssim eV_\text{sd}$
($\abs{\Delta_{10,00}} \lesssim eV_\text{sd}$), the bias on the drive QD opens
the $01\leftrightarrow 11$ ($10\leftrightarrow 00$) transition via sequential
tunneling, and the drag changes to cotunneling-assisted drag. This results in an
enhanced drag current compared to the cotunneling-only drag.

Figures~\ref{fig:drag_vertex}(b) and~\ref{fig:drag_vertex}(d) show the bias
dependence of the drive and drag currents at the gate voltages marked by dots in
Figs.~\ref{fig:drag_vertex}(a) and~\ref{fig:drag_vertex}(c). In the linear
low-bias regime, $eV_\text{sd} < k_BT$, $I_\text{drive}\propto V_\text{sd}$ and
$I_\text{drag} \propto V_\text{sd}^2$ for $\abs{\Delta_{10,01}}<k_BT$ (red,
yellow and green dots). The drag current is linear in $V_\text{sd}$ only at bias
voltages $eV_\text{sd}\ll k_BT$ (not shown). For $\abs{\Delta_{10,01}}> k_BT$
(blue dot), nonlocal cotunneling is exponentially suppressed,
$\Gamma_{10,01}\sim e^{-\Delta_{10,01}/k_BT}$, resulting in a vanishing drag
current. The drive current, however, remains finite due to elastic
cotunneling. In the nonlinear regime, $eV_\text{sd} > k_BT$,
$I_\text{drag} \sim V_\text{sd}^2$ up to
$eV_\text{sd}\sim\max (2\abs{\Delta_{10,01}},\abs{\Delta_{10/01,11}})$ where it
experiences a crossover to a $I_\text{drag} \sim \log V_\text{sd}$ dependence in
agreement with Eq.~\eqref{eq:drag_triple}. At even higher bias,
$eV_\text{sd} \gtrsim U_{12}$, the conventional drag
mechanism~\cite{Buttiker:Mesoscopic} which is driven by sequential tunneling
takes over (see also below).

\begin{figure}[!b]
  % \includegraphics[width=0.4\linewidth]{energydiagram}
  % \hspace{5pt}
  % \includegraphics[width=0.49\linewidth]{drag_sign}
  \includegraphics[width=.98\linewidth]{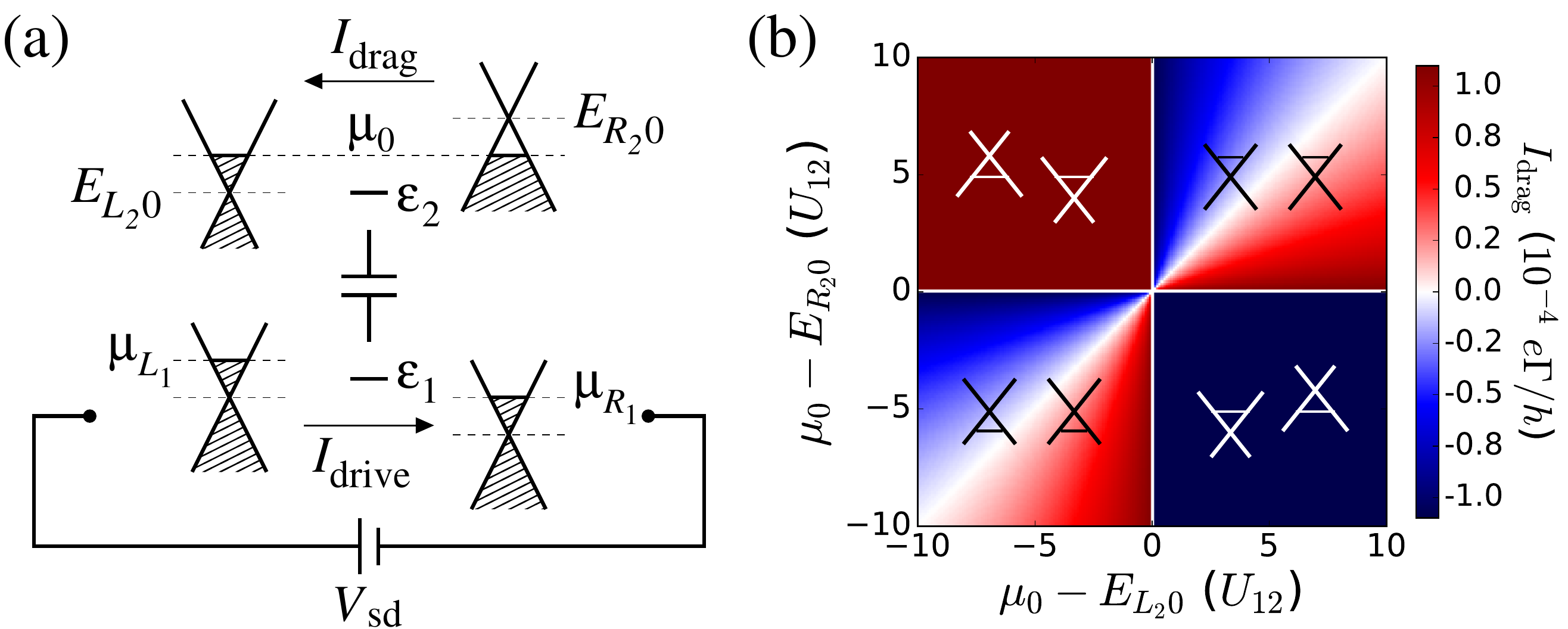}
  \caption{(a) Energy level diagram of the graphene-based CQD in
    Fig.~\ref{fig:overview}(c). The QD levels, $\varepsilon_i=-eV_i$, and the
    positions of the Dirac points in the leads, $E_{\alpha 0} = -e V_\alpha$,
    are controlled by local gates. (b) Drag current as a function of gate
    voltage on the leads of the drag system (see Dirac cone insets) at the upper
    triple point in the stability diagram. Parameters (in units of $U_{12}$):
    $U_{12} = 1$, $\Gamma_{L_10/R_10} = 0.01\equiv\Gamma$,
    $\Gamma_{L_20/R_20} \propto \abs{\mu_0 - E_{L_20/R_20}}$,
    $\partial\Gamma_{L_2/R_2} = \text{sgn}(\mu_{0}-E_{L_20/R_20})$,
    $t_{L_2}=t_{R_2}$, $eV_\text{sd}=0.1$, $k_BT=0.01$.}
\label{fig:drag_sign}
\end{figure}
From Eq.~\eqref{eq:drag_triple} it is clear that the direction of the drag
current depends, in a nontrivial way, on the lead couplings in the drag
system. This is demonstrated in Fig.~\ref{fig:drag_sign} which shows the drag
current at the upper triple point as a function of the positions of the Dirac
points in the drag leads. At the diagonal we have
$\Gamma_{L_2}(\varepsilon) = \Gamma_{R_2}(\varepsilon)$, and hence the drag
vanishes. Off the diagonal, $\Gamma_{L_2 0} \neq \Gamma_{R_2 0}$ and
$\partial\Gamma_{L_2} = \partial\Gamma_{R_2}$, the factor
$\Gamma_{L_2 0} - \Gamma_{R_2 0}$ governs the sign of the drag current. Upon
crossing the Dirac point in one of the leads, the drag changes sign due to an
inversion in the sign of the corresponding DOS derivative. Remarkably, the drag
becomes independent on $\abs{\mu_{0}-E_{L_20/R_20}}$ in this case. This follows
from the fact that for symmetric tunnel couplings
$\partial\Gamma_{L_2} = -\partial\Gamma_{R_2}$ which leads to a cancellation of
the $\Gamma_{L_2 0} + \Gamma_{R_2 0}$ factors in Eq.~\eqref{eq:drag_triple}. For
asymmetric tunnel couplings this is not the case. The unconventional sign of the
mesoscopic drag, which we have verified also holds for the conventional drag
mechanism~\cite{Buttiker:Mesoscopic}, is in stark contrast to that of the drag
in coupled graphene layers~\cite{Ponomarenko:Strong}.

\begin{figure}[!t]
  \centering
  % \includegraphics[scale=0.22]{stability_drive_current_vs_bias}
  % \hspace{0.5cm}
  % \includegraphics[scale=0.22]{stability_drive_current_vs_bias_log}
  % \includegraphics[scale=0.22]{stability_drag_current_vs_bias}
  % \hspace{0.5cm}
  % \includegraphics[scale=0.22]{stability_drag_current_vs_bias_log}
  \includegraphics[width=1.\linewidth]{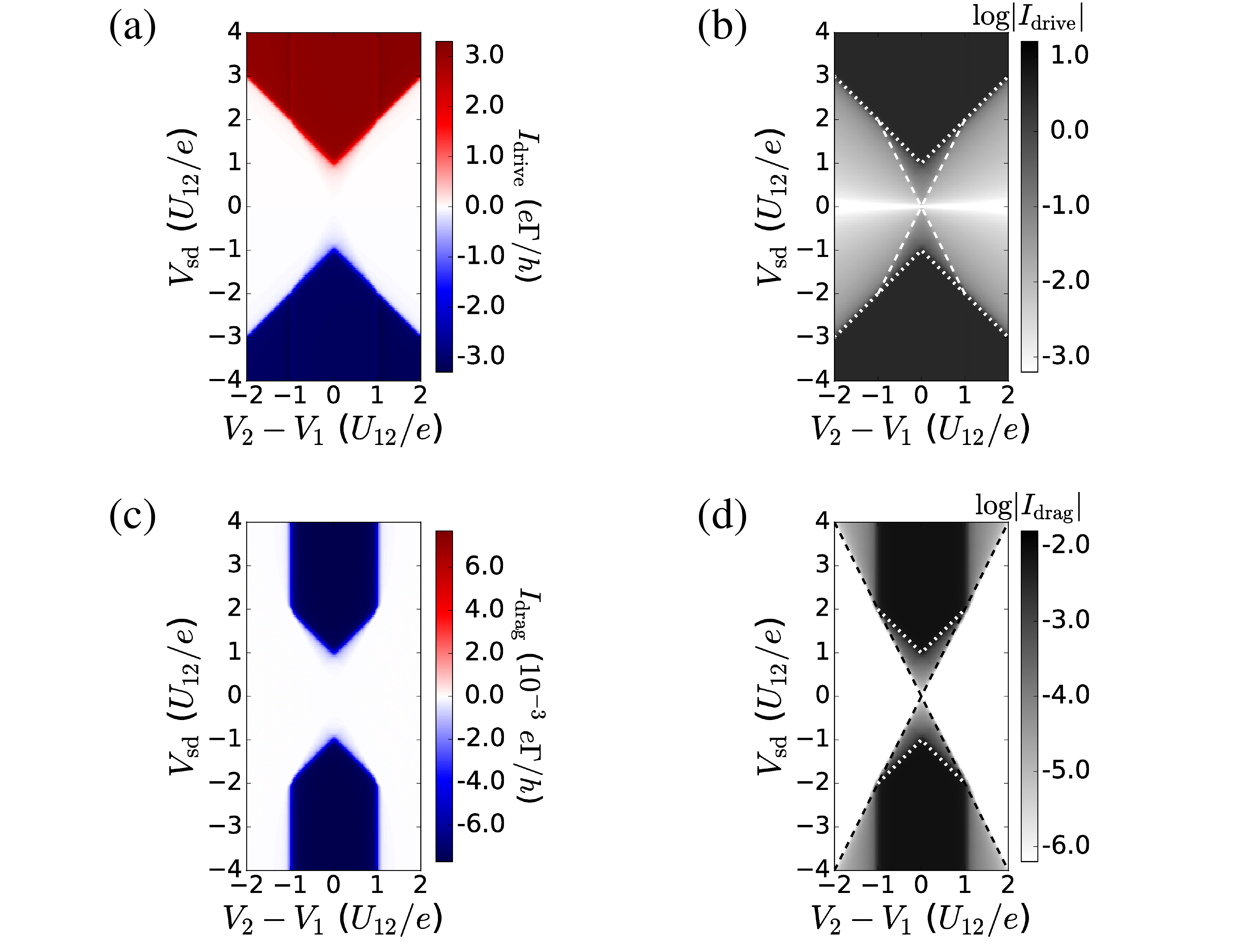}
  \caption{Bias spectroscopy. The plots show the current through drive (top) and
    drag (bottom) QDs at the center of the 10,01 degeneracy line with the bias
    applied to the drive QD. (a),(c): Linear scale. (b),(d): Log scale. The
    dashed (dotted) lines mark the boundaries to the regions where the currents
    are dominated by nonlocal cotunneling (sequential tunneling). See
    Fig.~\ref{fig:drag_vertex} for parameters.}
\label{fig:drag_bias}
\end{figure}

In the bias spectroscopy of the CQDs shown in Fig.~\ref{fig:drag_bias}, distinct
fingerprints of nonlocal cotunneling and the drag mechanism can be observed
inside the so-called Coulomb-blockade diamonds where the sequential tunneling
drive and drag currents are suppressed. It shows the drive (top) and drag
(bottom) currents at the center of the stability diagram (green dot in
Fig.~\ref{fig:drag_vertex}) as a function of gate detuning and drive bias. In
the low-bias Coulomb-blockaded regime,
$e\abs{V_\text{sd}}<\min(U_{12} + e\abs{V_2-V_1},2U_{12})$, nonlocal cotunneling
manifests itself in nonzero drive and drag currents in the region
$\abs{V_\text{sd}}/2 > \abs{V_2-V_1}$ which at $\Delta_{10,01}=0$ extends down
to zero bias.

At high bias, $e\abs{V_\text{sd}} > U_{12} + e\abs{V_2-V_1}$, sequential
tunneling dominates both the drive and drag currents. However, for
$e\abs{V_2-V_1} > U_{12}$ where the conventional drag
mechanism~\cite{Buttiker:Mesoscopic} is suppressed, cotunneling-assisted drag
extends the region with nonzero drag to $e\abs{V_2-V_1} < e\abs{V_\text{sd}}/2$.
The different slopes $s$ of the boundaries to the regions where, respectively,
sequential tunneling (dotted, $\abs{s}=1$) and nonlocal cotunneling (dashed,
$\abs{s}=2$) dominate the drive and drag currents (see log plots in
Fig.~\ref{fig:drag_bias}), is a direct fingerprint of nonlocal cotunneling and
its associated drag mechanism~\cite{supplementary}.

Finally, we estimate the magnitude of the drag current and comment on its
experimental verification. Taking $\Gamma_\alpha, k_B T \sim 0.1U_{12}$, a drag
current of the order of $I_\text{drag} \gtrsim (U_{12}/\mathrm{meV})^2$~pA is
predicted for the cotunneling-assisted drag at
$eV_\text{sd}\gtrsim k_BT , \max
(2\abs{\Delta_{10,01}},\abs{\Delta_{10/01,00/11}})$.
This is well within experimentally detectable currents and allows for a unique
identification of the nonlocal cotunneling-driven drag via its distinct
identifiers---i.e., the bias dependence in Eq.~\eqref{eq:drag_triple} and its
fingerprints in bias spectroscopy [Fig.~\ref{fig:drag_bias}(d)]. While the
high-bias cotunneling broadening of the drag region in
Fig.~\ref{fig:drag_bias}(d) was recently observed in
Ref.~\onlinecite{Ensslin:Measurement}, the drag at low bias remains unexplored.

\textbf{\emph{Conclusions.}}---In summary, we have identified a ratchetlike
drag mechanism in CQDs driven by nonlocal cotunneling processes. The key
ingredient for the drag mechanism is that the coupling to the leads be energy
dependent. This can be achieved via, e.g., gate-dependent tunnel
barriers~\cite{Ilani:Realization,Molenkamp:Three}, or be an intrinsic property
like in graphene-based QD structures with built-in graphene
leads~\cite{Ensslin:Measurement}. Studying the Coulomb drag in an idealized
version of such a QD structure, we demonstrated its nontrivial dependence on the
lead couplings and identified its fingerprints in bias spectroscopy. Possible
routes for future explorations of drag in CQDs include shot noise and cross
correlations
characteristics~\cite{Gossard:Tunable,Buttiker:Mesoscopic,Belzig:FCSShot}, the
effect of level broadening~\cite{Schon:Cotunneling,Romito:Measuring} and Kondo
physics~\cite{Gordon:Pseudospin,Sun:Orbital} which become important at
$\Gamma_\alpha > k_BT$, as well as drag due to other coupling mechanisms between
the QDs~\cite{Guo:Coupling}.

\begin{acknowledgments}
  \textbf{\emph{Acknowledgements.}}---We would like to thank J.~Santos and
  N.~A.~Mortensen for fruitful discussions, and K.~Ensslin and D.~Bischoff for
  clarifications on the experimental details in
  Ref.~\onlinecite{Ensslin:Measurement} and comments on the manuscript. The
  Center for Nanostructured Graphene (CNG) is sponsored by the Danish Research
  Foundation, Project DNRF103.
\end{acknowledgments}

\emph{Note added}.---While this work was under review, we became aware of a
related experimental work in which evidence of the nonlocal cotunneling drag
mechanism was observed at low bias in bias
spectroscopy~\cite{Gordon:Cotunneling}.

\bibliographystyle{apsrev}
\bibliography{paper.bbl}

\pagebreak
\widetext
\clearpage


\section*{Supplementary material}

# Supplemental material for "Correlated Coulomb drag in capacitively coupled quantum-dot structures"

Kristen Kaasbjerg* and Antti-Pekka Jauho
*Center for Nanostructured Graphene (CNG), Department of Micro- and Nanotechnology,
Technical University of Denmark, DK-2800 Kgs. Lyngby, Denmark*

(Dated: February 9, 2016)

## S1. BIAS SPECTROSCOPY WITH THE BIAS VOLTAGE APPLIED ASYMMETRICALLY

In experiments, the bias voltage on the drive system is often applied to the source, or the drain, electrode only. In order to ease the comparison with our theoretically calculated bias spectroscopy diagrams (Fig. 4 of the main text where the bias has been applied symmetrically to the source and drain electrodes) we give in Fig. S1 the results for an asymmetrically applied bias voltage, i.e. $\mu_{L_1} = eV_{\rm sd} + \mu_0$ and $\mu_{R_1} = \mu_0$.

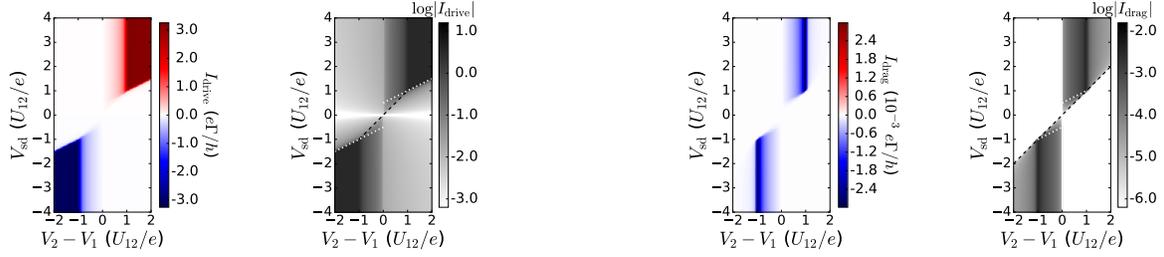

FIG. S1. (Color online) Current through the drive (left) and drag (right) dots at the center of the 10/01 degeneracy line vs bias and gate detuning with the bias on the drive dot applied asymmetrically, i.e. $\mu_{L_1} = eV_{\rm sd}$ and $\mu_{R_1} = 0$. Parameters (in units of $U_{12}$): $U_{12} = 1$, $\Gamma_{L_10/R_10} = \Gamma_{L_20/R_20} = 0.01 \equiv \Gamma$, $\partial \Gamma_{L_2} = -\partial \Gamma_{R_2}$, $t_{L_2} = t_{R_2}$, $k_B T = 0.01$.

The main difference between the bias spectroscopy diagrams in Fig. 4 of the main text and those in Fig. S1 above, is the suppression of the currents at negative (positive) detuning, $V_2 - V_1$, and positive (negative) bias polarity. Furthermore, the slopes $s$ of the boundaries to the regions where, respectively, sequential tunneling (dotted, $|s| = 1/2$) and nonlocal cotunneling (dashed, $|s| = 1$) dominates the drive and drag currents (see log plots in Fig. S1) are different.

Interestingly, we note that a feature similar to the one in the low-bias cotunneling-dominated drag current in the rightmost plot of Fig. S1, has been observed in the transconductance of two capacitively coupled QDs in Ref. 1 [their Fig. 4(g)].

## S2. COTUNNELING REGULARIZATION SCHEME

In this section, we generalize the commonly applied cotunneling regularization scheme[2,3] to the situation where the lead couplings are energy dependent. The result obtained here applies to general energy-dependent lead couplings which may originate from the either the lead density of states (DOS) and/or the tunnel couplings to the leads (see below). Furthermore, the generic form for the cotunneling rates considered below, allows for a straight-forward generalization to cotunneling in other QD systems.

From a $T$-matrix calculation[4] of cotunneling in the QD system considered in the main text, the rate for transferring an electron from lead $\alpha$ to lead $\beta$ in a cotunneling process (*elastic* or *nonlocal*) which, at the same time, changes the QD state from $|m\rangle$ to $|n\rangle$ ($|m\rangle = |n\rangle$ for *elastic* cotunneling) can be written on the generic form

$$\Gamma_{mn}^{\alpha\beta} = \int \frac{d\varepsilon}{2\pi\hbar} \Gamma_\alpha(\varepsilon + \Delta_{mn}) \Gamma_\beta(\varepsilon) \left| \frac{A}{\varepsilon - \Delta_1} + \frac{B}{\varepsilon - \Delta_2} \right|^2 f_\alpha(\varepsilon + \Delta_{mn})[1 - f_\beta(\varepsilon)]$$
$$= n_B(\mu_\beta - \mu_\alpha + \Delta_{mn}) \int \frac{d\varepsilon}{2\pi\hbar} \Gamma_\alpha(\varepsilon + \Delta_{mn}) \Gamma_\beta(\varepsilon) \left| \frac{A}{\varepsilon - \Delta_1} + \frac{B}{\varepsilon - \Delta_2} \right|^2 [f_\beta(\varepsilon) - f_\alpha(\varepsilon + \Delta_{mn})]. \quad (S1)$$

Here, $\Delta_{1,2}$ denotes the energy differences associated with the two possible intermediate states and $A, B = \pm 1$ for electron- and hole-like intermediate states, respectively. Furthermore, $\Gamma_\alpha(\varepsilon) = 2\pi\rho_\alpha(\varepsilon)|t_\alpha(\varepsilon)|^2$ is the *energy-dependent* lead coupling, $\rho_\alpha$ is the density of states (DOS) in lead $\alpha$, $t_\alpha$ is the tunneling amplitude, $\Delta_{mn} = E_n - E_m$ is the

energy difference between the initial and final states, $f_\alpha$ is the Fermi function of lead $\alpha$ with chemical potential $\mu_\alpha$, and $n_B$ is the Bose-Einstein distribution. In the last equality, we have recast the integrand into a form that is easier to tackle.

In the expression for the cotunneling rate above, the energy denominators associated with the virtually-occupied intermediate states give rise to a diverging cotunneling rate when the intermediate states go "on-shell". To deal with this divergence, we follow the standard regularization scheme[2,3] and introduce a phenomenological tunnel broadening of the QD states (not accounted for in a $T$-matrix calculation) by adding an imaginary infinitesimal $i\gamma$ (regularizer) in the energy denominators. The integral can then be evaluated after which the regularized cotunneling rates are obtained by taking the limit $\gamma \to 0$.

### A. General expression for the regularized cotunneling rates

With the regularizer added in the energy denominators of Eq. (S1), the cotunneling rate can be expressed as

$$\Gamma^{\alpha\beta}_{mn} = \frac{n_B(\mu_\beta - \mu_\alpha + \Delta_{mn})}{2\pi\hbar} \lim_{\gamma \to 0} \left[ A^2 I_1(\Delta_1) + B^2 I_1(\Delta_2) + 2AB I_2(\Delta_1, \Delta_2) \right] \tag{S2}$$

where the integrals $I_{1/2}$ are given by (with their $\Delta_n$ dependence suppressed in the following)

$$I_n = \int_{-\infty}^{\infty} d\varepsilon\, \Gamma_\alpha(\varepsilon + \Delta_{mn}) \Gamma_\beta(\varepsilon) F_n(\varepsilon) \left[ f_\beta(\varepsilon) - f_\alpha(\varepsilon + \Delta_{mn}) \right], \tag{S3}$$

and the functions $F_n$ are defined by the terms which result from the absolute-value squared factor,

$$\left| \frac{A}{\varepsilon - \Delta_1 + i\gamma} + \frac{B}{\varepsilon - \Delta_2 + i\gamma} \right|^2 = \left| \frac{A}{\varepsilon - \Delta_1 + i\gamma} \right|^2 + \left| \frac{B}{\varepsilon - \Delta_2 + i\gamma} \right|^2 + 2\mathrm{Re}\left( \frac{A}{\varepsilon - \Delta_1 + i\gamma} \frac{B}{\varepsilon - \Delta_2 - i\gamma} \right)$$
$$\equiv A^2 F_1(\varepsilon; \Delta_1) + B^2 F_1(\varepsilon; \Delta_2) + 2AB F_2(\varepsilon; \Delta_1, \Delta_2). \tag{S4}$$

This leaves us with two types of integrals, $I_{1,2}$, to evaluate.

We proceed to evaluate the integrals in Eq. (S3) using contour integration. To this end, we start by rewriting the Fermi functions as

$$f_\alpha(\varepsilon) = \frac{1}{2} \left[ 1 - \tanh\left( \frac{\varepsilon - \mu_\alpha}{2k_B T} \right) \right] = \frac{1}{2} \left[ 1 + \frac{i}{\pi} \left[ \Psi^+_\alpha(\varepsilon) - \Psi^-_\alpha(\varepsilon) \right] \right], \tag{S5}$$

where

$$\Psi^\pm_\alpha(\varepsilon) = \Psi\left( 1/2 \pm i\frac{\beta}{2\pi}(\varepsilon - \mu_\alpha) \right), \tag{S6}$$

$\Psi$ is the digamma function[5], and $\beta = 1/k_B T$. The two digamma functions $\Psi^\pm$ have poles $z_n = \mu_\alpha \pm i 2\pi/\beta \left(n + \frac{1}{2}\right)$, $n \in \mathbb{N}$, which lie in the positive and negative complex half-planes, respectively. We therefore split up the integral into two subintegrals, $I_n = I^+_n - I^-_n$, which deal with the contributions from $\Psi^\pm$ separately,

$$I^\pm_n = \frac{i}{2\pi} \int_{-\infty}^{\infty} d\varepsilon\, \Gamma_\alpha(\varepsilon + \Delta_{mn}) \Gamma_\beta(\varepsilon) F_n(\varepsilon) \left[ \Psi^\pm_\beta(\varepsilon) - \Psi^\pm_\alpha(\varepsilon + \Delta_{mn}) \right]. \tag{S7}$$

This has the advantage that upon extending the integrations to closed contours in the complex plane, the poles of the $\Psi^\pm$ functions can be avoided by closing the contours in the half plane where $\Psi^\pm$ do not have any poles. If the lead couplings $\Gamma_{\alpha/\beta}$ are well-behaved functions without singularities, the resulting contour integrals $I^\pm_c$ are thus given by the sum over residues of the integrand at the poles $z_i$ of $F_n$ which are enclosed by the integration contours,

$$I^\pm_{cn} = \frac{i}{2\pi} \oint_{C_\mp} dz\, \Gamma_\alpha(z + \Delta_{mn}) \Gamma_\beta(z) F_n(z) \left[ \Psi^\pm_\beta(z) - \Psi^\pm_\alpha(z + \Delta_{mn}) \right] = 2\pi i \sum_i \mathrm{Res}(z_i). \tag{S8}$$

Here, the contours are chosen as semicircles $C_\mp$ with radius $R \to \infty$ closed in the negative/positive complex half-planes, respectively (see Fig. S2).

Writing the contour integrals as a sum over the different contributions from the integration contour,

$$I^\pm_{cn} = \mp I^\pm_n + I^\pm_{C_{R_\mp} n}, \tag{S9}$$

where $\mp$ in front of the first term accounts for the direction of the contour along the real axis and $I^\pm_{C_{R_\mp}}$ denote the contribution from the semi-circle arcs, the integrals along the real axis in Eq. (S3) can be obtained as

$$I_n = I_n^+ - I_n^- = I^+_{C_{R_-}n} + I^-_{C_{R_+}n} - I^+_{cn} - I^-_{cn} \equiv I_{C_R n} - I_{cn}. \tag{S10}$$

In the following subsections, we evaluate the different contributions to the $I_{1/2}$ integrals one by one. The contributions from the semi-circle arcs $C_{R_\pm}$ need special attention since they, depending on the energy dependence of the lead couplings, might have nonzero values.

### 1. $I_{cn}$: Contributions from residues

As the integrands in Eq. (S8) are of the form $g(z)/h(z)$, where $h$ denotes the denominators of the $F_n$ functions defined in Eq. (S4), the residues at the simple poles $z_i$ of $F_n$, where $h(z_i) = 0$, $h'(z_i) = 0$ and $g(z_i) \neq 0$, can be obtained as $\text{Res}(g/h, z_i) = g(z_i)/h'(z_i)$.

$$I_{c1}: F_1(z) = \frac{1}{(z-\Delta_i)^2 + \gamma^2}$$

The function $F_1$ has simple poles in $z_i^\pm = \Delta_i \pm i\gamma$, $i = 1$ or $i = 2$, which contribute to the integrals $I_{ci}^\mp$, respectively. From the residues we get

$$I_{c1}^\pm = \pm 2\pi i \frac{i}{2\pi} \frac{i}{2\gamma} \Gamma_\alpha(\Delta_i + \Delta_{mn} \mp i\gamma)\Gamma_\beta(\Delta_i \mp i\gamma)\left[\Psi_\beta^\pm(\Delta_i \mp i\gamma) - \Psi_\alpha^\pm(\Delta_i + \Delta_{mn} \mp i\gamma)\right] \tag{S11}$$

The sum of the contributions is

$$-I_{c1} = -I_{c1}^+ - I_{c1}^- = \frac{i}{2\gamma}\bigg[\Gamma_\alpha(\Delta_i + \Delta_{mn} + i\gamma)\Gamma_\beta(\Delta_i + i\gamma)\left[\Psi_\alpha^-(\Delta_i + \Delta_{mn} + i\gamma) - \Psi_\beta^-(\Delta_i + i\gamma)\right]$$

$$+ \Gamma_\alpha(\Delta_i + \Delta_{mn} + i\gamma)^*\Gamma_\beta(\Delta_i + i\gamma)^*\left[\Psi_\beta^-(\Delta_i + i\gamma)^* - \Psi_\alpha^-(\Delta_i + \Delta_{mn} + i\gamma)^*\right]\bigg]$$

$$= \frac{1}{\gamma}\text{Im}\left[\Gamma_\alpha(\Delta_i + \Delta_{mn} + i\gamma)\Gamma_\beta(\Delta_i + i\gamma)\left[\Psi_\beta^-(\Delta_i + i\gamma) - \Psi_\alpha^-(\Delta_i + \Delta_{mn} + i\gamma)\right]\right] \tag{S12}$$

$$= -\frac{1}{\gamma}\text{Im}\left[\Gamma_\alpha(\Delta_i + \Delta_{mn} - i\gamma)\Gamma_\beta(\Delta_i - i\gamma)\left[\Psi_\beta^+(\Delta_i - i\gamma) - \Psi_\alpha^+(\Delta_i + \Delta_{mn} - i\gamma)\right]\right], \tag{S13}$$

where we have used the property $\Psi(z)^* = \Psi(z^*) \to \Psi^+(z^*) = \Psi^-(z)^*$ of the digamma function.

This contribution diverges when we take the $\gamma \to 0$ limit. Following the standard recipe[2,3], we therefore expand in $\gamma$, $I_{c1} \approx I_{c1}(\gamma=0) + \gamma I'_{c1}(\gamma=0)$, and discard the diverging $O(\gamma^{-1})$ term which can be associated with sequential tunneling already included in a separate lowest-order calculation of the tunneling rates. The remaining $O(\gamma^0)$ term gives the desired contribution to the cotunneling rate,

$$-I_{c1}[O(\gamma^0)] = \text{Re}\left[\left[\Gamma'_\alpha(\Delta_i + \Delta_{mn})\Gamma_\beta(\Delta_i) + \Gamma_\alpha(\Delta_i + \Delta_{mn})\Gamma'_\beta(\Delta_i)\right]\left[\Psi_\beta^-(\Delta_i) - \Psi_\alpha^-(\Delta_i + \Delta_{mn})\right]\right]$$

$$+ \frac{\beta}{2\pi}\text{Im}\left[\Gamma_\alpha(\Delta_i + \Delta_{mn})\Gamma_\beta(\Delta_i)\left[\Psi_\beta^{-'}(\Delta_i) - \Psi_\alpha^{-'}(\Delta_i + \Delta_{mn})\right]\right], \tag{S14}$$

where $\Gamma'$ denotes the derivative of the lead couplings and $\Psi'$ is the trigamma function.

$$I_{c2}: F_2(z) = \text{Re}\frac{1}{(z-\Delta_1+i\gamma)(z-\Delta_2-i\gamma)}$$

The function $F_2$ has simple poles in $z_i^\pm = \Delta_i \pm i\gamma$, $i = 1, 2$, which contribute to the integrals $I_{c2}^\mp$, respectively. From the residues we get

$$I_{c2}^+ = 2\pi i \frac{i}{2\pi} \frac{1}{2}\bigg[\frac{1}{\Delta_1 - \Delta_2 - 2i\gamma}\Gamma_\alpha(\Delta_1 + \Delta_{mn} - i\gamma)\Gamma_\beta(\Delta_1 - i\gamma)\left[\Psi_\beta^+(\Delta_1 - i\gamma) - \Psi_\alpha^+(\Delta_1 + \Delta_{mn} - i\gamma)\right]$$

$$+ \frac{1}{\Delta_2 - \Delta_1 - 2i\gamma}\Gamma_\alpha(\Delta_2 + \Delta_{mn} - i\gamma)\Gamma_\beta(\Delta_2 - i\gamma)\left[\Psi_\beta^+(\Delta_2 - i\gamma) - \Psi_\alpha^+(\Delta_2 + \Delta_{mn} - i\gamma)\right]\bigg] \tag{S15}$$

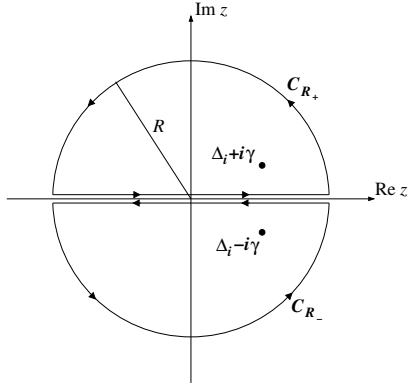

FIG. S2. Integration contours $C_\pm$ for the integrals $I_n^\mp$ in Eq. (S8). The contour integrals are given by the residues at the simples poles $z_i^\pm = \Delta_i \pm i\gamma$ of the $F_n$ functions defined in Eq. (S4).

and

$$I_{c2}^- = 2\pi i \frac{i}{2\pi} \frac{1}{2}\left[\frac{1}{\Delta_1 - \Delta_2 + 2i\gamma}\Gamma_\alpha(\Delta_1 + \Delta_{mn} + i\gamma)\Gamma_\beta(\Delta_1 + i\gamma)\bigl[\Psi_\beta^-(\Delta_1 + i\gamma) - \Psi_\alpha^-(\Delta_1 + \Delta_{mn} + i\gamma)\bigr]\right.$$
$$\left. + \frac{1}{\Delta_2 - \Delta_1 + 2i\gamma}\Gamma_\alpha(\Delta_2 + \Delta_{mn} + i\gamma)\Gamma_\beta(\Delta_2 + i\gamma)\bigl[\Psi_\beta^-(\Delta_2 + i\gamma) - \Psi_\alpha^-(\Delta_2 + \Delta_{mn} + i\gamma)\bigr]\right], \quad (S16)$$

respectively.

The sum of the contributions is

$$-I_{c2} = -I_{c2}^+ - I_{c2}^- = \text{Re}\frac{1}{\Delta_1 - \Delta_2 - 2i\gamma}\left[\Gamma_\alpha(\Delta_1 + \Delta_{mn} - i\gamma)\Gamma_\beta(\Delta_1 - i\gamma)\bigl[\Psi_\beta^+(\Delta_1 - i\gamma) - \Psi_\alpha^+(\Delta_1 + \Delta_{mn} - i\gamma)\bigr]\right.$$
$$\left. - \Gamma_\alpha(\Delta_2 + \Delta_{mn} + i\gamma)\Gamma_\beta(\Delta_2 + i\gamma)\bigl[\Psi_\beta^-(\Delta_2 + i\gamma) - \Psi_\alpha^-(\Delta_2 + \Delta_{mn} + i\gamma)\bigr]\right]$$
$$\stackrel{\gamma \to 0}{=} \frac{1}{\Delta_1 - \Delta_2}\text{Re}\left[\Gamma_\alpha(\Delta_1 + \Delta_{mn})\Gamma_\beta(\Delta_1)\bigl[\Psi_\beta^+(\Delta_1) - \Psi_\alpha^+(\Delta_1 + \Delta_{mn})\bigr]\right.$$
$$\left. - \Gamma_\alpha(\Delta_2 + \Delta_{mn})\Gamma_\beta(\Delta_2)\bigl[\Psi_\beta^-(\Delta_2) - \Psi_\alpha^-(\Delta_2 + \Delta_{mn})\bigr]\right]. \quad (S17)$$

Here we do not encounter any problems taking the $\gamma \to 0$ limit.

### 2. $I_{C_Rn}$: Contributions from semi-circle arcs

In order to calculate the contributions from the integrals along the semi-circle arcs in the limit $R \to \infty$, we use the asymptotic expansion of the digamma function[5]

$$\lim_{|z| \to \infty} \Psi(z) = -\gamma + \sum_{n=1}^\infty \frac{z}{n(n+z)} = \ln z - \frac{1}{2z} - \sum_{n=1}^\infty \frac{B_{2n}}{2nz^{2n}}, \quad (S18)$$

where $\gamma$ is the Euler-Mascheroni constant and $B_{2n}$ are Bernoulli numbers. From the asymptotic expansion, we find for the factors in Eq. (S8) involving the $\Psi^\pm$ functions

$$\lim_{|z|\to\infty}\left[\Psi_\beta^+(z) - \Psi_\alpha^+(z + \Delta_{mn})\right] = \ln\left[1 + \frac{i\beta}{2\pi}\frac{\mu_\alpha - \mu_\beta - \Delta_{mn}}{1/2 + i\beta/2\pi(z + \Delta_{mn} - \mu_\alpha)}\right] - \frac{1}{1 + i\beta(z - \mu_\beta)/\pi} + \frac{1}{1 + i\beta(z + \Delta_{mn} - \mu_\alpha)/\pi} + \ldots$$
$$\approx \ln\left[1 + \frac{\mu_\alpha - \mu_\beta - \Delta_{mn}}{z}\right] + i\frac{\mu_\beta - \mu_\alpha + \Delta_{mn}}{\beta z^2/\pi} + \ldots$$
$$\approx \frac{\mu_\alpha - \mu_\beta - \Delta_{mn}}{z} \quad (S19)$$

and

$$\lim_{|z|\to\infty} \left[\Psi_\beta^-(z) - \Psi_\alpha^-(z+\Delta_{mn})\right] = \ln\left[1 - \frac{i\beta}{2\pi}\frac{\mu_\alpha - \mu_\beta - \Delta_{mn}}{1/2 - i\beta/2\pi(z+\Delta_{mn}-\mu_\alpha)}\right] - \frac{1}{1 - i\beta(z-\mu_\beta)/\pi} + \frac{1}{1 - i\beta(z+\Delta_{mn}-\mu_\alpha)/\pi} + \ldots$$
$$\approx \ln\left[1 + \frac{\mu_\alpha - \mu_\beta - \Delta_{mn}}{z}\right] - i\frac{\mu_\alpha - \mu_\beta - \Delta_{mn}}{\beta z^2/\pi} + \ldots$$
$$\approx \frac{\mu_\alpha - \mu_\beta - \Delta_{mn}}{z}, \tag{S20}$$

respectively.

In addition we have that $\lim_{|z|\to\infty} F_n(z) = z^{-2}$. Inserting in Eq. (S8) with $z = Re^{i\theta} \to dz = iRe^{i\theta}d\theta$ on the semi-circle arcs, we find

$$\begin{aligned}
I_{C_R n} &= I_{C_{R_-}}^+ + I_{C_{R_+}}^- \\
&= \frac{i^2}{2\pi}\int_{-\pi}^0 d\theta\, Re^{i\theta}\Gamma_\alpha(Re^{i\theta})\Gamma_\beta(Re^{i\theta})\frac{1}{R^2e^{2i\theta}}\frac{\mu_\alpha - \mu_\beta - \Delta_{mn}}{Re^{i\theta}} \\
&+ \frac{i^2}{2\pi}\int_0^\pi d\theta\, Re^{i\theta}\Gamma_\alpha(Re^{i\theta})\Gamma_\beta(Re^{i\theta})\frac{1}{R^2e^{2i\theta}}\frac{\mu_\alpha - \mu_\beta - \Delta_{mn}}{Re^{i\theta}} \\
&= \frac{\mu_\beta - \mu_\alpha + \Delta_{mn}}{2\pi}\int_{-\pi}^\pi d\theta\,\frac{\Gamma_\alpha(Re^{i\theta})\Gamma_\beta(Re^{i\theta})}{R^2 e^{2i\theta}}
\end{aligned} \tag{S21}$$

Thus, as long as the asymptotic behavior of the lead couplings is $\lim_{|z|\to\infty}\Gamma_\alpha(z) = z^n$ with $n < 1$, the integrals along the semi-circle arcs vanish for $R \to \infty$. For graphene leads, this is not the case (see below).

It should be noted that for $A = -B$, i.e. with one electron- and one hole-like intermediate state, the semi-circle contributions from the different terms in Eq. (S2) cancel in the cotunneling rate.

### B. Linearized lead couplings

At low bias (compared to the energy scale at which the lead couplings show nonlinear energy dependence), we can expand the lead couplings around the equilibrium chemical potentials $\mu_0$ of the leads, $\Gamma_\alpha(\varepsilon) \approx \Gamma_{\alpha 0} + \xi\partial\Gamma_\alpha$, where $\xi = \varepsilon - \mu_0$, $\Gamma_{\alpha 0} = \Gamma_\alpha(\mu_0)$, and $\partial\Gamma_\alpha = \partial\Gamma_\alpha/\partial\varepsilon|_{\varepsilon=\mu_0}$. With this approximation, $\lim_{|z|\to\infty}\Gamma_\alpha(z) = z\partial\Gamma_\alpha$, and the integrals along the semi-circle arcs in Eq. (S21) attain a finite value,

$$I_{C_R n} = \partial\Gamma_\alpha \partial\Gamma_\beta \left(\mu_\beta - \mu_\alpha + \Delta_{mn}\right). \tag{S22}$$

It should be pointed out that the apparent contradiction associated with the fact that we here have used the $|z| \to \infty$ limit of an expansion valid at low energies only, is both physically and mathematically sound. The integral along the real axis in Eq. (S1) [and Eq. (S7)] is cut off by the Fermi functions and does *not* depend on the behavior of the lead couplings at $\varepsilon \to \infty$. Only when the real-axis integrals in Eq. (S7) are expressed in terms of the different contributions to the contour integrals in Eq. (S8), as in Eq. (S10), does the $z \to \infty$ limit of the lead couplings enter.

*Graphene leads*

For bulk graphene leads with linear DOS, $\rho(\varepsilon) \propto |\varepsilon|$, the lead couplings acquire the linear energy dependence of the DOS if the tunnel couplings are assumed independent on energy, i.e. $\Gamma(\varepsilon) \propto \rho(\varepsilon)$. In this case the result in Eq. (S22) above applies. However, it should be noted that, as the graphene DOS is nonanalytic at the Dirac point where the DOS vanishes, the result obtained here strictly only holds away from the Dirac point.

---



\end{document}